\documentclass[%
aip,jcp,amsmath,amssymb, reprint, twocolumn
]{revtex4-2}
\usepackage{amsmath}
\usepackage{mathtools}
\usepackage{float}
\usepackage{amsfonts}
\usepackage{amssymb}
\usepackage{dcolumn} 
\usepackage{array}
\newcolumntype{P}[1]{>{\centering\arraybackslash}p{#1}}
\newcolumntype{M}[1]{>{\centering\arraybackslash}m{#1}}
\newcolumntype{C}[1]{>{\centering\arraybackslwash}p{#1}}
\usepackage{float}
\usepackage{psfrag}
\usepackage{tabularx}
\usepackage{stackengine}
\usepackage{amssymb}
\usepackage{mathtools}  
\usepackage{xfrac} 
\usepackage[T1]{fontenc}
\usepackage{graphicx}
\usepackage{diagbox}
\usepackage[caption=false]{subfig}
\usepackage{tikz}
\usetikzlibrary{positioning}
\usetikzlibrary{arrows}
\usetikzlibrary{trees}
\usepackage{amssymb}
\usetikzlibrary{decorations.pathmorphing}
\usetikzlibrary{decorations.markings}
\usetikzlibrary{automata,positioning}
\usepackage{braket}
\usepackage{rotating}
\usepackage{adjustbox}
\usepackage{ragged2e}
\usepackage{simplewick}
\usepackage{simpler-wick}

\usepackage{csquotes}
\usepackage{physics,amsmath}
\usepackage{xcolor}
\usepackage{fancyhdr}
\UseRawInputEncoding
\usepackage[normalem]{ulem}

\usepackage{scalerel}
\usepackage{hyperref}
\usepackage{appendix}

\begin{document}

\author{Chayan Patra}
\affiliation{ Department of Chemistry,  \\ Indian Institute of Technology Bombay, \\ Powai, Mumbai 400076, India}

\author{Debaarjun Mukherjee}
\affiliation{ Department of Chemistry,  \\ Indian Institute of Technology Bombay, \\ Powai, Mumbai 400076, India}

\author{Sonaldeep Halder}
\affiliation{ Department of Chemistry,  \\ Indian Institute of Technology Bombay, \\ Powai, Mumbai 400076, India}

\author{Dibyendu Mondal}
\affiliation{ Department of Chemistry,  \\ Indian Institute of Technology Bombay, \\ Powai, Mumbai 400076, India}

\author{Rahul Maitra}
\email{rmaitra@chem.iitb.ac.in}
\affiliation{ Department of Chemistry,  \\ Indian Institute of Technology Bombay, \\ Powai, Mumbai 400076, India}
\affiliation{Centre of Excellence in Quantum Information, Computing, Science \& Technology, \\ Indian Institute of Technology Bombay, \\ Powai, Mumbai 400076, India}
\title{Towards a Resource-Optimized Dynamic Quantum Algorithm via Non-iterative Auxiliary Subspace Corrections}

\begin{abstract}

Recent quantum algorithms pertaining to electronic structure theory primarily focus on threshold-based dynamic construction of ansatz by selectively including important many-body operators.
These methods can be made systematically more accurate by tuning the threshold
to include more number of operators into the ansatz.
However, such improvements come at the cost of rapid proliferation of the circuit depth,
especially for highly correlated molecular systems.
In this work, we address this issue by
the development of a novel theoretical framework that
relies on the segregation of an ansatz into a dynamically selected core \enquote{\textit{principal}} component,
which is, by construction adiabatically decoupled from the remaining operators.
This enables us to perform computations involving the \textit{principal} component using extremely shallow-depth circuits
whereas, the effect of the remaining \enquote{\textit{auxiliary}} component
is folded into the energy function via a cost-efficient non-iterative correction, ensuring the
requisite accuracy.
We propose a formalism that analytically predicts the \textit{auxiliary}
parameters from the \textit{principal} ones, followed by a suite of non-iterative
\textit{auxiliary subspace correction} techniques with different levels of sophistication.
The \textit{auxiliary subspace corrections} incur no additional quantum resources, yet complement
an inadequately expressive core of the ansatz to recover significant amount of electronic correlations.
We have numerically validated the resource efficiency
and accuracy of our formalism with a number of strongly correlated molecular systems. 
\end{abstract}

\maketitle

\section{Introduction}

Quantum computing algorithms have recently garnered significant attention as they show immense promise to solve certain
classically intractable many-body problems.
Quantum computers leverage the
principles of superposition and entanglement
to bypass exponential or higher order polynomial scaling that
existing classical electronic structure methods
\cite{white1992density,huron1973iterative,buenker1974individualized,cc3,cc4,cc5,bartlett2007coupled,crawford2000introduction}
suffer from.
Quantum phase estimation (QPE) \cite{abrams1997simulation,abrams1999quantum}
was one of the first algorithms to theoretically showcase the quantum advantage.
However, it suffers from deep quantum circuits, which results in huge accumulation of noise and extremely erroneous results
when implemented using the currently available noisy intermediate scale quantum (NISQ) computers\cite{preskill2018quantum}.
To bypass this, hybrid quantum-classical algorithms like variational quantum eigensolver (VQE)\cite{peruzzo2014variational}
in the unitary coupled cluster (UCC) framework have recently gained immense popularity as they work by evaluating expectation
values of parameterized operators in a quantum processing unit while the parameter optimization is outsourced to classical optimizers.

Another class of algorithm
known as the projective quantum eigensolver (PQE)\cite{stair2021simulating} takes a different approach by optimizing the parameters using coupled cluster-like residual
constructions. PQE can converge to identical energy values as that of VQE with similar or fewer quantum resources. It also shows inherent resilience to hardware noise.
These algorithms can be applied to simulate chemical systems in tandem with several error mitigation strategies\cite{temme2017error,song2019quantum,zhang2020error,koczor2021exponential,huggins2021virtual,cai2023quantum} to combat
different sources of hardware errors
for practical applicability. 
However, the scalability of VQE or PQE for dealing with larger molecular systems is restricted due to hardware limitations of NISQ devices and
fundamental constraints of the error mitigation
strategies\cite{takagi2022fundamental}. This necessitates the development of algorithms that can reduce the resource requirements of a given ansatz with reduced circuit complexity for less impact of noise
while maintaining the characteristic accuracy as that of the original ansatz. Along this line, several algorithms have recently been developed
in both VQE and PQE framework such as measurement-based iterative ansatz compactification methods \cite{grimsley2019adaptive,delgado2021variational,mondal2023development,feniou2023overlap,zhao2023orbital, tang2021qubit, ostaszewski2021structure, tkachenko2021correlation, zhang2021adaptive, sim2021adaptive,halder2022dual,halder2023corrections},
machine-learning based resource-efficient methods\cite{halder2023machine,liu2023training,sonaldeep_rbm,mehendale2023exploring},
qubit-space based CNOT reduction techniques\cite{ryabinkin2018qubit,ryabinkin2020iterative,yordanov2020efficient,magoulas2023cnot},
effective-Hamiltonian driven downfolding methods \cite{kowalski2021dimensionality,kowalski2023quantum,huang2023leveraging,bauman2019downfolding,metcalf2020resource},
methods of moments coupled cluster (MMCC\cite{piecuch1994solving,kowalski2000method}) inspired quantum algorithms \cite{magoulas2023unitary,peng2022mapping,claudino2021improving} 
and measurement-free dimensionality reduction via many-body perturbative methods\cite{patra2024projective,halder2024noise}
to mention some of them.
Most of these methods rely upon a threshold based selection of a reduced subspace containing \enquote{important} operators.
Properly tuned inclusion conditions on the threshold can enhance accuracy but lead to more complex circuits, which are often impractical for molecular applications in the NISQ era.
Contrarily,
a somewhat less stringent inclusion criteria compromises accuracy by
selecting fewer important operators although it results in shallower circuits. Owing to the current hardware limitations,
such shallower circuits are desired even though the corresponding compact ansatz is often
not expressive enough for accurate energy calculations since it is attributed to only a relatively small number 
operators.

In this work, we address this problem by presenting a theoretical formalism that
enables one to recover accurate electronic correlations
even with such inadequately expressive dynamic ansatz.
In this study, we have considered selected PQE (SPQE)\cite{stair2021simulating} as the dynamic selection protocol,
which performs quantum measurements and iteratively picks up the important operators based on a threshold.
Here we have shown that SPQE implicitly presumes
a hierarchical structure among the parameters according to their convergence behavior
which conforms to the \textit{adiabatic approximation}\cite{patra2024projective,patra2023synergistic,agarawal2021approximate,halder2023development},
previously introduced by some of the present authors.
It ensures a decoupling of the parameter space into a dominant, slower converging \textit{principal parameter subset (PPS)}
which is dynamically selected and optimized via SPQE.
The leftover submissive, faster-converging parameters are collectively called the \textit{auxiliary parameter subset (APS)}.
Owing to this adiabatic decoupling, the corresponding operator space can be divided into a \textit{principal} unitary subspace
which forms the core of the ansatz and an \textit{auxiliary} subspace.
This particular structure of the unitary allows us to define an Adiabatically Decoupled UCC (AD-UCC) ansatz, which plays a central role in the entire analysis.
We have further developed a mathematical framework that predicts the leftover \textit{auxiliary} parameters from the SPQE-driven \textit{principal} parameters.
This inter-relationship between the \textit{principal} and \textit{auxiliary} parameters is established
based on our recently developed \textit{slaving principle} in the 
framework of nonlinear electronic structure optimization
problems in both classical\cite{patra2023synergistic,agarawal2021approximate,agarawal2021accelerating,agarawal2022hybrid}
and quantum\cite{patra2024projective,halder2023development} computing parlance.
Further, the effect of the \textit{auxiliary} parameters
is folded into the energy functional in an approximate manner via non-iterative \textit{auxiliary subspace corrections} (ASC)
without the need for any additional circuit resources.
The advantage of our method is particularly more pronounced with less accurate core ansatz
where ASC effectively complements the core for a dramatic improvement of the energy accuracy
at an extremely reduced circuit depth.
Since the algorithm enforces adiabatic decoupling through SPQE with 
a posteriori ASC,
we refer to this algorithm as AD(SPQE)-ASC.
Though we have considered SPQE framework here for the analysis, adiabatic decoupling
is a general formalism and can be extended to other dynamic algorithms as well.
To numerically establish the efficacy of our method, we have
studied the relative performances of SPQE and AD(SPQE)-ASC with different levels of non-iterative approximations
for challenging molecular cases such as stretching of linear $H_4$, 
linear $H_6$ and symmetric stretching of $O-H$ bond in $H_2O$ which are considered to 
be prototypical models for electronic strong correlation effects.

The manuscript is structured as follows, we start with the preliminary discussion on PQE and SPQE. Then we move over
to the mathematical formalism of AD(SPQE)-ASC which contains efficient \textit{principal} parameter selection,
predicting the \textit{auxiliary} parameters from the \textit{principal} ones and designing different variants
of post-optimization non-iterative corrections to the energy function. Subsequently, we present our numerical results
to establish the superiority of the algorithm.

\section{Theory}
\subsection{Selected Projective Quantum Eigensolver and the Spontaneous Emergence of Adiabatic Decoupling Conditions} \label{section for spqe}

The ground state of a molecular Hamiltonian can be approximated by disentangled unitary coupled cluster\cite{evangelista2019exact} (dUCC) ansatz
$\ket{\Psi}=\hat{U}(\boldsymbol{\theta})\ket{\Phi_0}$ parameterized by $\boldsymbol{\theta}$, where $\ket{\Phi_0}$ is the Hartree-Fock (HF)
reference state and $\hat{U}$ is taken to be
\begin{equation} \label{disentangled unitary op}
    \hat{U}(\boldsymbol{\theta}) = \prod_{\mu} e^{\theta_{\mu}\hat{\kappa}_{\mu}}
\end{equation}
where, $\hat{\kappa}_{\mu}=\hat{\tau}_{\mu}-\hat{\tau}_{\mu}^\dagger$ is an anti-Hermitian operator with $\hat{\tau}_{\mu}= \hat{a}^{\dagger}_{a}\hat{a}^{\dagger}_{b}....\hat{a}_{j}\hat{a}_{i}$ and the boldface $\boldsymbol{\theta}$ represents the parameter vector.
Here, $\mu$ represents a multi-index particle-hole excitation structure as defined by the string of creation ($\hat{a}^{\dagger}$) and annihilation ($\hat{a}$) operators with the indices $\{i,j,...\}$ denoting the occupied spin-orbitals in the HF state and $\{a,b,...\}$ denoting the unoccupied spin-orbitals.
Since all the unitary operators are parameterized, we will interchangeably use the words \enquote{operator} and \enquote{parameters} depending on the context.
In PQE\cite{stair2021simulating} framework the parameters that satisfy
the Schrodinger equation can be obtained by iterative optimization via $\theta_{\mu}^{(m+1)} = \theta_{\mu}^{(m)} + \frac{r_{\mu}^{(m)}}{D_{\mu}}$
where, $m$ is the iterative step counter and $D_\mu$ is the M{\o}ller-Plesset denominator. Here, $r_\mu$ is the residual element constructed by taking projections against excited determinants $\ket{\Phi_\mu}$ 
\begin{equation} \label{r_mu}
   r_{\mu}(\boldsymbol{\theta}) = \bra{\Phi_{\mu}}\hat{U}^{\dagger}(\boldsymbol{\theta}) \hat{H} \hat{U}(\boldsymbol{\theta})\ket{\Phi_o} ; \mu \ne 0.
\end{equation}
The residuals can be efficiently calculated using a quantum computer as they can be expressed as a sum of
diagonal quantities\cite{stair2021simulating}
\begin{equation} \label{r_mu with diagonal quantities}
    r_{\mu} = \bra{\Omega_{\mu}(\sfrac{\pi}{4})} \bar{H} \ket{\Omega_{\mu}(\sfrac{\pi}{4})} - \frac{1}{2}E_{\mu} - \frac{1}{2}E_0
\end{equation}
with the following definitions, $\ket{\Omega_{\mu}(\theta)} = e^{\hat{\kappa}_{\mu} \theta} \ket{\Phi_0}$, $\bar{H} = \hat{U}^{\dagger}\hat{H}\hat{U}$, $E_\mu = \bra{\Phi_\mu} \bar{H} \ket{\Phi_\mu}$
and $E_0 = \bra{\Phi_0} \bar{H} \ket{\Phi_0}$.
Due to the exponential parameterization of the ansatz and the non-terminating nature of the BCH
expansion stemming from the non-commutativity of the anti-Hermitian $\hat{\kappa}_\mu$ operators, the optimization is highly coupled and nonlinear in nature.
In such an iterative optimization problem, one may implicitly introduce the notion of discrete-time
which may be assumed to be embedded in each iteration step,
as often done in studying nonlinear \textit{iterative maps}\cite{strogatz2018nonlinear}. Thus, this iteration step
\enquote{$m$} is to be taken as discrete time step
such that the variation of the parameters in two subsequent iterations can be expressed as
\begin{equation} \label{delta theta}
    \Delta \theta^{(m)}_\mu = \frac{r_\mu^{(m)}}{D_\mu}
\end{equation}
where, $\Delta \theta^{(m)}_\mu=\theta^{(m+1)}_\mu-\theta^{(m)}_\mu$ with $m$ being the iteration number.
The iterative procedure is deemed converged when the residual condition
\begin{equation} \label{residual condition}
    r_\mu \rightarrow 0
\end{equation}
is satisfied.
For PQE with $n-$tuple excitation operators, 
the iterative procedure requires the
construction and optimization of $(n_on_v)^n$ residue elements
which often leads to deep quantum circuits.
A more lucrative alternative is to dynamically select only a handful 
of operators that are important (such that the gate depth can be kept 
low) for a specific molecule and 
specific nuclear arrangement at the cost of additional quantum 
measurements. It can be done with selected PQE (SPQE)\cite{stair2021simulating}
which starts with an operator pool and consists of two sets of iterative cycles, namely 
\textit{macro-iterations} for selecting the important operators from the pool and \textit{micro-iterations} to optimize the corresponding parameters.
At $k$-th macro-iteration,
a \textit{residual state} defined as
\begin{equation} \label{residual state}
\begin{split}
    \ket{\Tilde{r}^{(k)}} &= (1 + i\Delta{t}\hat{U}^{\dagger (k)}\hat{H}\hat{U}^{(k)})) \ket{\Phi_0} \\
    & = C_0^{(k)} \ket{\Phi_0} + \sum_\mu C_\mu^{(k)} \ket{\Phi_\mu}
\end{split}
\end{equation}
is constructed, where $\hat{U}^{(k)}$ consists of of all the selected operators in the previous $0$ to $(k-1)$ macro-iterations.
The coefficients $C_\mu^{(k)}$ are proportional to the residuals $r_\mu^{(k)}=\hat{U}^{\dagger (k)}\hat{H}\hat{U}^{(k)}$
and given by approximately\cite{stair2021simulating} $\mid C_\mu^{(k)} \mid \approx \Delta t \mid r_\mu^{(k)} \mid$
where $\Delta t$ is the evolution time which is in general small in magnitude.
The residual state is measured to get a probability distribution of the excited determinants $\ket{\Phi_\mu}$. From the entire operator pool, those operators are excluded
that have the summation of their residual squared less than the square of a pre-defined macro-iteration threshold $\Omega$, such that
\begin{equation} \label{operator selection criteria}
    \sum_\mu^{excluded} \mid r_\mu^{(k)} \mid^2 \hspace{2mm} \approx \sum_\mu^{excluded} \frac{\mid C_\mu^{(k)} \mid^2}{\Delta t^2} \le \Omega^2 \mbox{ (exclusion criteria)}    
\end{equation}
The remaining excitation operators are then selected at $k$-th macro-iteration, optimized using PQE micro-iteration cycles (which uses Eq. \eqref{delta theta}) and appended to $\hat{U}^{(k)}$ for generating the residual state
in the next $(k+1)$-th iteration via Eq. \eqref{residual state} till all the unique operators to be added are exhausted. Thus, using an alternate combination of macro and micro iterations, SPQE can construct a dUCC ansatz sequence consisting of only ``important'' operators.

The discussions so far suggest that the selection of operators in SPQE is entirely dependent upon the macro-iteration
threshold ($\Omega$). SPQE can be made systematically more accurate by tuning $\Omega$ to include more operators into the ansatz,
however, at the cost of increased circuit depth.
To circumvent this, we introduce a mathematical framework here requiring minimal gate-depth SPQE computations on a quantum processor
that can cater to the needs of modern NISQ devices. In addition to that, a set of non-iterative energy corrections complements for the lost correlation,
ensuring the requisite accuracy of the problem.
Towards this development, we borrow some of the
ideas from our recent works\cite{patra2024projective,halder2023development} that establishes  
that there exists an explicit hierarchy regarding the \textit{convergence timescale}
(or, number of iterations required for convergence) of the parameters which
enables an adiabatic decoupling of the entire operator space into a
dominant \textit{principal} and a submissive \textit{auxiliary} subpart
having different timescale of equilibration.
We argue and demonstrate that a problem-specific and dynamically flexible \textit{principal}-\textit{auxiliary} decoupling, which has profound mathematical
consequences towards efficiency and efficacy (\textit{vide infra}),
spontaneously emerges from SPQE as follows:
\begin{itemize}
    \item It is evident from Eq. \eqref{delta theta} that the parameter variation is solely dictated by
the magnitude of the associated residual
\begin{equation} \label{proportionality condition}
    \Delta \theta_{\mu} \propto r_\mu.
\end{equation}
Since operator exclusion/inclusion criteria (Eq. \eqref{operator selection criteria}) explicitly depends upon the absolute magnitude of the residuals $r_{\mu}$,
this proportionality condition (Eq. \eqref{proportionality condition}) immediately ensures that the parameters selected via SPQE
are usually those that take much more time (iteration steps) to converge (in a conventional PQE) owing to the larger magnitude of the associated residual.
These parameters span the \textit{PPS} and are denoted as ($\theta_P^{(SPQE)}$) since 
this subset is constructed
via SPQE measurements. The number of elements in this space is $N_P$. Due to their importance from a many-body perspective, this subspace form the core of the bipartitely decoupled ansatz. 
    \item The complementary operator pool that satisfies the exclusion criteria (Eq. \eqref{operator selection criteria}), are not selected in the core, owing to its smaller magnitude of the residuals and consequently faster timescale of equilibration.
    The associated parameters span the \textit{APS} with the corresponding number of elements denoted by $N_A$. Therefore, the whole parameter set 
    ($\{\theta\}$) can now be divided into \textit{principal} and \textit{auxiliary} parameters such that:
\begin{equation}
    \{\theta\} = \{\theta_P^{(SPQE)}\} \oplus \{\theta_A\}
\end{equation}
The smaller relative magnitude of the residual elements for the excluded parameters
ensures that the
absolute magnitude of the \textit{principal} parameters are significantly larger
than that of the \textit{auxiliary} ones
\begin{equation} \label{parameter amplitude condition}
    \mid \theta_P^{(SPQE)} \mid \hspace{2mm} >> \hspace{2mm} \mid \theta_A \mid.
\end{equation}
\end{itemize}
To translate this notion of parameter
space segregation into the operator space, we can define a \textit{principal-auxiliary bipartite unitary} operator ($\hat{U}_{pab}$) such that
\begin{equation} \label{U_pab principal auxiliary bipartite op}
    \hat{U}_{pab} = \hat{U}_P(\boldsymbol{\theta_P}^{(SPQE)}) \cdot \hat{U}_A(\boldsymbol{\theta_A})
\end{equation}
where, $\hat{U}_P(\boldsymbol{\theta_P}^{(SPQE)})$ and $\hat{U}_A(\boldsymbol{\theta_A})$ are \textit{principal} and \textit{auxiliary} unitary operators, respectively,
each possessing the disentangled form as defined in
Eq. \eqref{disentangled unitary op}.
For notational convenience
we will henceforth denote $\boldsymbol{\theta_P}^{(SPQE)}$ as $\boldsymbol{\theta_P}$.
This particular structure of the unitary leads to the following form of the associated correlated state
\begin{equation}
    \ket{\Psi_{AD-UCC}} = \hat{U}_{pab}\ket{\Phi_0} = e^{\hat{\kappa}_P \theta_P} e^{\hat{\kappa}_A \theta_A} \ket{\Phi_0} 
\end{equation}
which we term as the Adiabatically Decoupled UCC (AD-UCC) ansatz.
The ordering of $\hat{U}_{pab}$ is such that the Hamiltonian transformation begins with the largest
many-body rotation and continues in a descending manner
\begin{equation} \label{SPQE operator ordering}
\begin{split}
    H & \rightarrow \hat{U}_A^{\dagger} \cdot \hat{U}_P^{\dagger} \hat{H} \hat{U}_P \cdot \hat{U}_A\\
    & = . . .e^{-\theta_{A_{\mu_2}} \hat{\kappa}_{A_{\mu_2}}} e^{-\theta_{A_{\mu_1}} \hat{\kappa}_{A_{\mu_1}}}. . .e^{-\theta_{P_{\mu_2}} \hat{\kappa}_{P_{\mu_2}}}e^{-\theta_{P_{\mu_1}}\hat{\kappa}_{P_{\mu_1}}} \\
    & \hat{H} e^{\theta_{P_{\mu_1}} \hat{\kappa}_{P_{\mu_1}}} e^{\theta_{P_{\mu_2}} \hat{\kappa}_{P_{\mu_2}}}. . .e^{\theta_{A_{\mu_1}} \hat{\kappa}_{A_{\mu_1}}} e^{\theta_{A_{\mu_2}} \hat{\kappa}_{A_{\mu_2}}}. . .   
\end{split}
\end{equation}
where, $\mid \theta_{P_{\mu_1}} \mid > \mid \theta_{P_{\mu_2}} \mid . . . > \mid \theta_{A_{\mu_1}} \mid > \mid \theta_{A_{\mu_2}} \mid ...$ and so on
with subscript $P$ $(A)$ denoting the elements of the \textit{principal} (\textit{auxiliary}) subspace.
One may of course design an energy function where the \textit{auxiliary} parameters are neglected altogether,
which is referred to as SPQE energy function
\begin{equation} \label{spqe energy}
    E_{SPQE} = \bra{\Phi_0} \hat{U}_P(\boldsymbol{\theta_P}) \hat{H} \hat{U}_P(\boldsymbol{\theta_P}) \ket{\Phi_0}.
\end{equation}
Such an ansatz where only $\theta_P$s are included can be made arbitrarily accurate as mentioned earlier by tuning $\Omega$ to
exclude less and involve more number of operators (and parameters) into $\hat{U}_P$.
This, however, leads to a rapid proliferation of the associated circuit depth.
Therefore, in the NISQ era, it is desirable to develop algorithms that can maintain the requisite accuracy even with
extremely compact core ansatz
to improve scalability. 
In such a scenario, the \textit{auxiliary} counterpart $\hat{U}_A$, which is completely ignored in the above expression (Eq. \eqref{spqe energy}), can have important
contributions to the correlation energy specially in the strongly correlated regime.
With an in-built recipe for adiabatic decoupling by SPQE, in the next section we develop
a framework that provides the posteriori energy correction to factor in the \textit{auxiliary} operators without any additional
major quantum resources.


\subsection{Synergistic Mapping of the Auxiliary Parameters and Non-iterative Auxiliary Subspace Corrections to Energy} \label{synergistic mapping}

The discussions of section \ref{section for spqe} suggest that \textit{principal} parameters obtained from SPQE with
appropriate conditions on $\Omega$ are pivotal for accurate energy predictions of a molecular system.
However, for a resource-efficient implementation of SPQE, we need to judiciously incorporate the larger \textit{auxiliary} subspace
into the energy function. This is required to complement the inadequately parameterized compact ansatz and retrieve the electronic correlation.
Towards this, we can invoke the \textit{adiabatic approximation}\cite{patra2024projective,patra2023synergistic,agarawal2021approximate,halder2023development,haken1983nonlinear,wunderlin1981generalized}
which allows us to neglect the variation of \textit{auxiliary} parameters in successive iterations
\begin{equation}
    \Delta \theta_{A_\alpha}=0 \hspace{3mm} \mbox{     (adiabatic approximation)}
\end{equation}
since it converges significantly faster than the \textit{principal} parameters.
Along with this approximation, the condition in Eq. \eqref{parameter amplitude condition}
can be utilized to design a mathematical function that predicts the
\textit{auxiliary} parameters from the information regarding the optimized \textit{principal} parameters only.
This is executed via computing the residual-like matrix elements of the similarity transformed Hamiltonian $\hat{U}_P^{\dagger}\hat{H}\hat{U}_P$
projected against the \textit{auxiliary}-space excited determinants
(see Appendix \ref{auxiliary mapping} for detailed steps)
\begin{equation} \label{theta_A as a function of theta_P}
    \begin{split}
        \theta_{A_\alpha} = \frac{1}{D_{A_\alpha}}\bra{\Phi_{A_\alpha}} \hat{U}_{P}^{\dagger}(\boldsymbol{\theta_P}) \hat{H} \hat{U}_{P}(\boldsymbol{\theta_P}) \ket{\Phi_0}
    \end{split}
\end{equation}
which can be efficiently implemented on a quantum computer by expanding it in terms of
three diagonal expectation values as shown in Eq. \eqref{r_mu with diagonal quantities}.
The expression for \textit{principal-to-auxiliary mapping}
in Eq. \eqref{theta_A as a function of theta_P} is a general equation
that gives us an accurate estimate of the \textit{auxiliary} 
parameters where the coupling between the \textit{principal} and \textit{auxiliary} parameters are
switched off during the optimization of the former.
In general this mapping is done as a single step procedure only after the \textit{principal} parameters are optimized.

However, the explicit computation of the matrix elements 
may potentially be bypassed particularly for SPQE-based adiabatic decoupling.
This is due to the fact that at the last macro-iteration ($k=k_{max}$) cycle $\hat{U}^{(k_{max})}=\hat{U}_P$ (see Eq. \eqref{residual state}, \eqref{operator selection criteria}) and the 
matrix element in Eq. \eqref{theta_A as a function of theta_P} is equivalent to those $r_\mu$'s which remain excluded. One may of course 
construct the matrix elements explicitly upon the dynamic selection of the
\textit{principal} parameters using any method of choice other than SPQE, which may incur nominal measurement overhead
as will be discussed shortly.
Further, with the criteria that the \textit{auxiliary} parameters are in general smaller in magnitude (Eq. \eqref{parameter amplitude condition}),
the energy function can be approximated in a way that contains the effect of larger many-body rotations
via \textit{principal} parameters exactly and folds in the effect of the \textit{auxiliary} parameters
approximately as correction terms with the second-order truncation of the
Baker-Campbell-Hausdorff (BCH) expansion
\begin{equation} \label{energy expression}
\begin{split}
     E & = \bra{\Phi_0}\hat{U}_A^\dagger \hat{U}_P^\dagger \hat{H} \hat{U}_P \hat{U}_A \ket{\Phi_0}\\
     & \approx \underbrace{\bra{\Phi_{0}} \bar{H}_P  \ket{\Phi_{0}}}_{{E_{SPQE}}} + \underbrace{\sum_{A_\alpha \in APS} \theta_{A_\alpha} (\boldsymbol{\theta_P}) \bra{\Phi_{0}}  [\bar{H}_P,\hat{\kappa}_{A_\alpha}] \ket{\Phi_{0}}}_{\mbox{Term 1}}\\
    & +\underbrace{\frac{1}{2} \sum_{A_\alpha,A_\beta} \theta_{A_\alpha}(\boldsymbol{\theta_P}) \theta_{A_\beta}(\boldsymbol{\theta_P}) \bra{\Phi_{0}} \Big[[\Bar{H}_P,\hat{\kappa}_{A_\alpha}],\hat{\kappa}_{A_\beta}\Big] \ket{\Phi_{0}}}_{\mbox{Term 2}}
\end{split}                                                                                                                                                                                                  \end{equation}
where, $\bar{H}_P = \hat{U}^{\dagger}_P \hat{H} \hat{U}_P$.
Note that, in this context the particular structure of the \textit{principal-auxiliary bipartite} operator $\hat{U}_{pab}$ (Eq. \eqref{U_pab principal auxiliary bipartite op})
is extremely important in attaining the approximate energy function here.
In the above expression \textit{Term 1} and \textit{Term 2} are only added after
the optimization, once we have access to the mapped \textit{auxiliary} parameters via Eq. \eqref{theta_A as a function of theta_P}.
Thus, these two terms are collectively referred to as the post-optimization non-iterative \textit{auxiliary subspace correction} (ASC)
terms. Computation of \textit{Term 1} on a quantum computer is relatively straightforward
as it can be evaluated analytically into the following expression (see Appendix \ref{appendix: energy equation derivation} for detailed derivation)
\begin{equation}
    \sum_{A_\alpha \in APS} \theta_{A_\alpha} (\boldsymbol{\theta_P}) \bra{\Phi_{0}}  [\bar{H}_P,\hat{\kappa}_{A_\alpha}] \ket{\Phi_{0}} = 2 \theta_{A_\alpha}^2 D_{A_\alpha}
\end{equation}
Evaluating \textit{Term 2} exactly is particularly non-trivial and since it already contains nonlinearity, the BCH expansion of the similarity transformed Hamiltonian can be
approximated to leading order bare Hamiltonian only, i.e. $\Bar{H}_P \approx \hat{H}$. 
We then invoke the M{\o}ller-Plesset partitioning of the Hamiltonian with canonical spin orbitals such that the Hamiltonian can be written as
$\hat{H}= \hat{F}^{(0)}+\hat{V}^{(1)}$, where $\hat{F}^{(0)}=\sum_q f_q \hat{a}^\dagger_q \hat{a}_q$ is the zeroth order one-body Fock operator
with $f_q$ being the one-body integrals
and $\hat{V}^{(1)}$ is the first order two-body operator. 
\textit{Term 2} can be approximated in two different ways
and we refer to these two approximations as scheme-I and scheme-II
(also represented as AD(SPQE)-ASC (I/II)). 
In scheme-I, only the zeroth order Fock operator is retained in the Hamiltonian and it leads to the following simplified energy expression
\begin{equation} \label{energy scheme I}
    E^{(I)} = \underbrace{\bra{\Phi_{0}} \hat{U}_P^{\dagger}(\boldsymbol{\theta_P}) \hat{H} \hat{U}_P (\boldsymbol{\theta_P})  \ket{\Phi_{0}}}_{E_{SPQE}} + \underbrace{\sum_{A_\alpha} \theta_{A_\alpha}^2 D_{A_\alpha}}_{ASC (I)}
\end{equation}
It is interesting to note that the ASC term for scheme-I, which spontaneosuly emerges from the \textit{adiabatic approximation}
and \textit{slaving principle}, shares some level of structural kinship to the non-iterative corrections
stemming from method of moments CC (MMCC) theory
\cite{piecuch2000search,kowalski2000method,kowalski2004new,pittner2009method,magoulas2023unitary}.
However, as one can see from the analysis so far,
the mathematical origin and underlying approximations for moments corrections and
ASC are fundamentally different.
On the other hand, in scheme-II the full Hamiltonian is retained and the energy
expression becomes
\begin{equation}
\begin{split} \label{energy scheme II}
    & E^{(II)} = \underbrace{\bra{\Phi_{0}} \hat{U}_P^{\dagger}(\boldsymbol{\theta_P}) \hat{H} \hat{U}_P (\boldsymbol{\theta_P})  \ket{\Phi_{0}}}_{E_{SPQE}} + \\
    & \underbrace{2 \sum_{A_\alpha} \theta_{A_\alpha}^2 D_{A_\alpha} + \frac{1}{2} \sum_{A_\alpha} \theta_{A_\alpha}^2 \bra{\Phi_{0}} \Big[[H,\hat{\kappa}_{A_\alpha}],\hat{\kappa}_{A_\alpha}\Big] \ket{\Phi_{0}}}_{ASC (II)}
\end{split}
\end{equation}
The detailed steps leading to the above expressions in Eq. \eqref{energy scheme I} and \eqref{energy scheme II} are shown in Appendix \ref{appendix: energy equation derivation}.
Thus, the benefit of this entire protocol is that one can obtain and optimize \textit{principal} parameters via SPQE with a compact circuit
while \textit{auxiliary} parameters and energy are calculated using
Eq. \eqref{theta_A as a function of theta_P}, \eqref{energy scheme I} (scheme-I) and \ref{energy scheme II} (scheme-II)
with the same shallow-circuit depth as that of SPQE since all the evaluations are constrained entirely to the \textit{principal} subspace.
The additional computations necessary for ASC in Eq. \eqref{energy scheme I} and \eqref{energy scheme II}
(both scheme-I and II) is governed
by the \textit{principal-to-auxiliary mapping} (Eq. \eqref{theta_A as a function of theta_P})
which in general requires $dim\{APS\}=N_A$ number of extra residual element evaluations
of the type $\bra{\Phi_\mu} \Bar{H} \ket{\Phi_0}$.
However, as discussed earlier in this section, for the particular case of SPQE, the mapping via Eq. \eqref{theta_A as a function of theta_P}
can be extracted from the SPQE selection protocol itself. Thus specifically for SPQE, ASC does not require any additional
residue evaluations which saves a significant amount of computational overhead.
Scheme-II further requires
some nominal amount of measurements stemming from the last term of Eq. \eqref{energy scheme II} for a better approximation.
Note that this extra measurements for scheme-II do not involve any circuits with unitary operators
and are significantly less than the overall SPQE measurement cost
(see Appendix \ref{measurement cost calculation} for details).
Since the adiabatic decoupling is obtained via SPQE and it includes a non-iterative ASC for energy, we term this formalism as
AD(SPQE)-ASC that consists of the following steps-
\begin{figure*}[!ht]
    \centering  
\includegraphics[width=\textwidth]{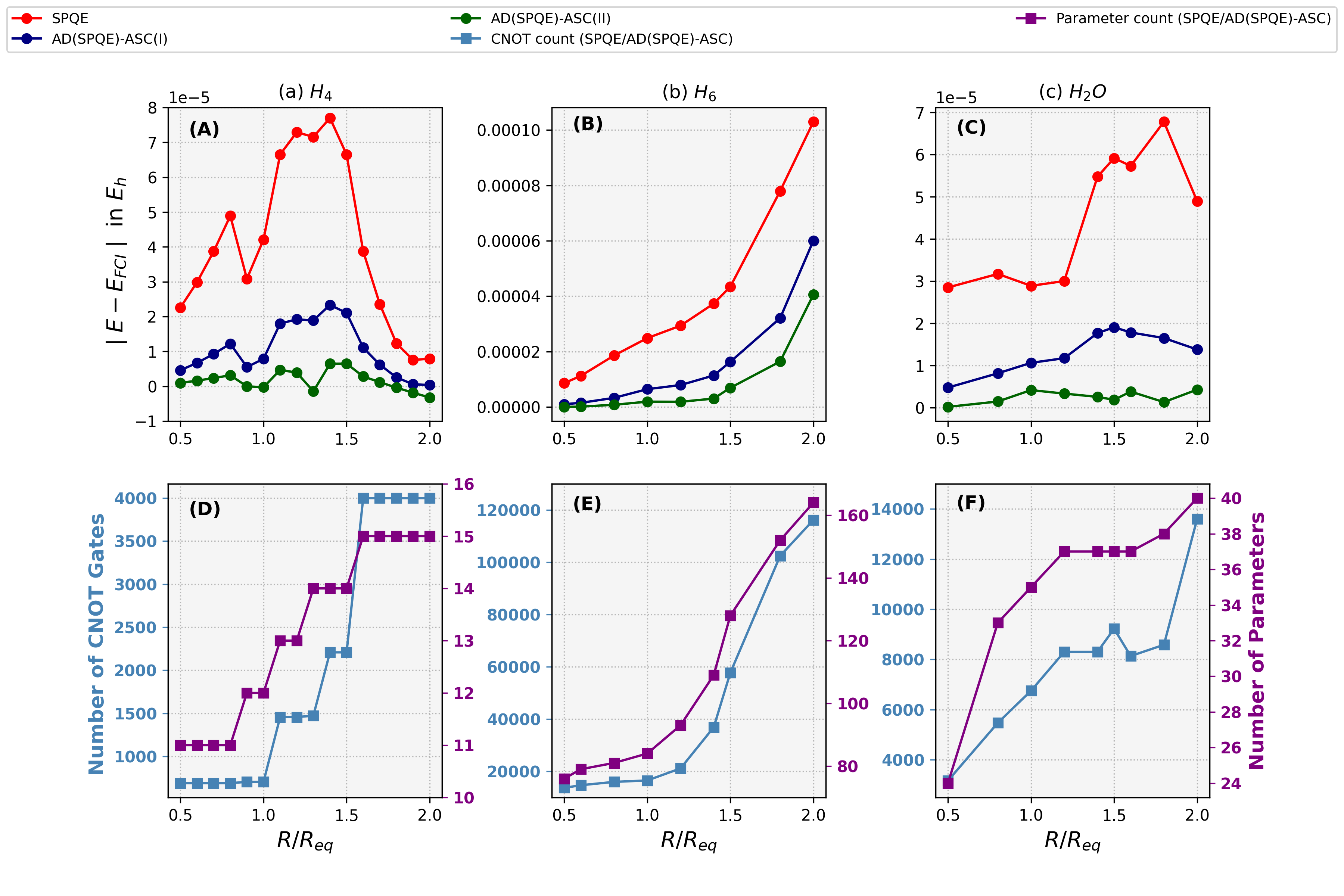}
\caption{Comparison between SPQE and AD(SPQE)-ASC (scheme-I and II) for bond stretching of linear (a) $H_4$ and (b) $H_6$ chains and (c) symmetric stretching of $H_2O$
with the information regarding the molecular geometries given in section \ref{results and discussion section}. SPQE
and AD(SPQE)-ASC contain up to quadruples excitation operators in the pool. The first row ((A), (B), (C)) represents the energy error with respect to FCI over the potential energy surface.
In the second row ((D), (E), (F))
we show the CNOT counts and parameter counts (in the twin y-axis) along various nuclear arrangements for SPQE and AD(SPQE)-ASC.
Note that for a given $\Omega$, both the methods require the same quantum resources.}
    \label{adspqe_pot_en_surf}
\end{figure*}

\begin{itemize}
    \item \textit{Principal Parameter Selection and Optimization}: SPQE is performed by the construction of residual
    state (Eq. \eqref{residual state}) at $k$-th macro-iteration and the important operators are filtered according to the condition in Eq. \eqref{operator selection criteria}.
    The corresponding parameters are optimized using micro-iteration cycles via Eq. \eqref{delta theta} and appended to $\hat{U}^{(k)}$ to form $\hat{U}^{(k+1)}$. This updated unitary is used to form the residual state at $(k+1)$-th step and the entire protocol involving macro and micro-iterations continues. When no new operators are added to the unitary, SPQE terminates and we have an ordered and optimized set of operators
    which can be taken to be the \textit{PPS}.

    \item \textit{Post-Optimization Principal-to-Auxiliary Mapping}: The leftover operators not taken into the ansatz by SPQE form the \textit{APS}. These \textit{auxiliary} parameters
    are mapped from the \textit{principal} parameters via the relationship established in Eq. \eqref{theta_A as a function of theta_P}.

    \item \textit{Non-iterative Correction to Energy}: The final energy is obtained using the expressions Eq. \eqref{energy scheme I} for scheme-I and Eq. \eqref{energy scheme II}
    for scheme-II.
\end{itemize}

In the next section we will discuss the accuracy and efficiency of this algorithm for challenging molecular applications.

\begin{figure*}[!ht]
    \centering  
\includegraphics[width=\textwidth]{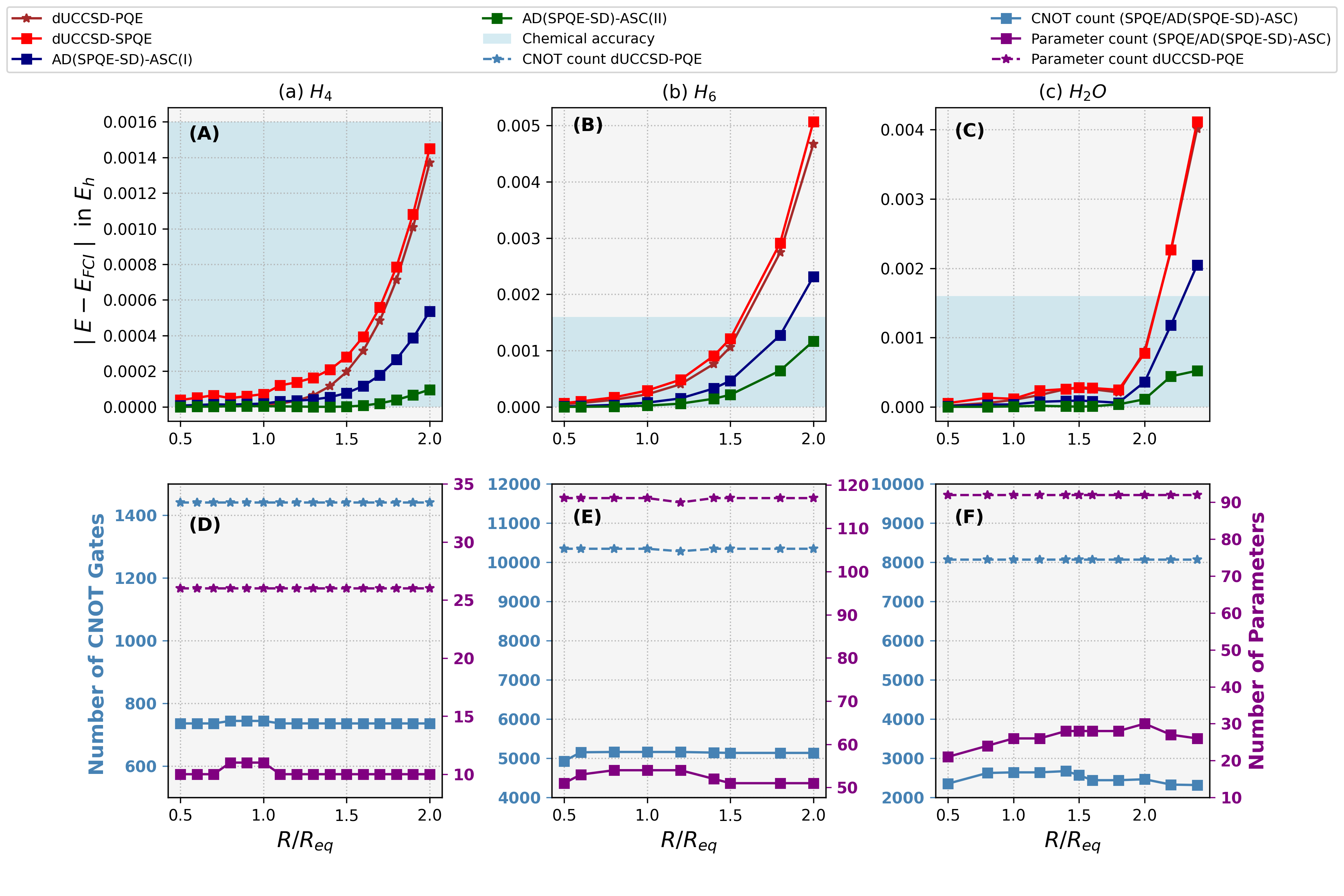}
\caption{Comparison between dUCCSD-SPQE and AD(SPQE-SD)-ASC (scheme-I and II) for bond stretching of linear (a) $H_4$ and (b) $H_6$ chains, and (c) symmetric stretching of $H_2O$. For comparison, the conventional dUCCSD-PQE results are also plotted.
First row ((A), (B), (C)) shows the energy difference from FCI against different geometrical arrangements of the molecules. The second row ((D), (E), (F)) shows CNOT and parameter counts for the corresponding
molecular configurations.}
    \label{adspqe_CCSD_TQ_pot_en_surf}
\end{figure*}

\section{Results and Discussion} \label{results and discussion section}
In this section we will illustrate the numerical results of our theoretical development by comparing AD(SPQE)-ASC (scheme I and II) with SPQE.
The numerical studies were performed with an in-house python code implemented by interfacing QForte\cite{stair2022qforte} and Qiskit\cite{qiskit2024}.
For the analysis we have studied the ground state potential energy surface (PES) for
(a) linear $H_4$ (equilibrium distance $R_{(eq)H-H}=0.75$\AA), (b) linear $H_6$ ($R_{(eq)H-H}=0.75$\AA) and (c) symmetric $O-H$ bond stretching
of $H_2O$ molecule ($R_{(eq)O-H}= 0.958$\AA, $\angle H-O-H =104.4776^\circ$ with frozen core approximation)
which are considered to be standard systems to study the electronic strong-correlation behaviour.
We have used STO-3G basis set for all our calculations with the electronic integrals imported from PySCF\cite{sun2018pyscf} driver in Qiskit and
from Psi4\cite{smith2020psi4} in QForte. The fermion-to-qubit mapping of the Hamiltonian is done using Jordan-Wigner\cite{jordan1993paulische} transformation.
For all our studies, SPQE operator selection and energy calculation is done by QForte with micro-iteration convergence threshold $10^{-5}$ and $\Delta t = 0.001$.
The selected excitation operators are then mapped to Qiskit convention for the computation of ASC.
All the methods under consideration contains singles, doubles, triples and upto quadruples (SDTQ) excitations
unless mentioned otherwise.

\begin{figure*}[!ht]
    \centering  
\includegraphics[width=\textwidth]{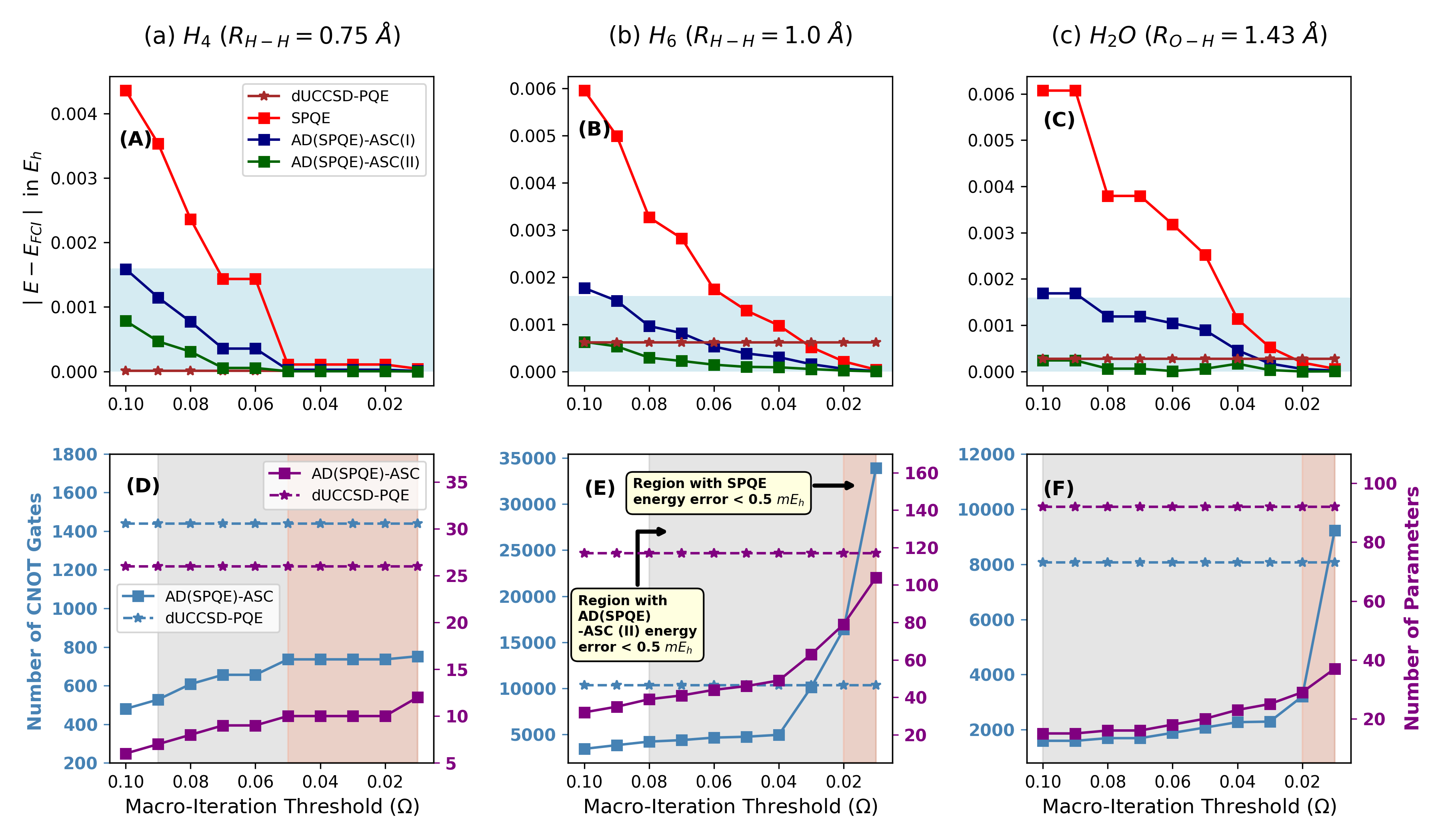}
\caption{Comparison between SPQE, AD(SPQE)-ASC (scheme-I and II) (with SDTQ pool) with different macro-iteration threshold ($\Omega$)
and dUCCSD-PQE for near equilibrium geometries of (a) $H_4$,
(b) $H_6$ and (c) $H_2O$
with the bond distances mentioned in each column. Here (A), (B) and (C) represent the absolute energy error with respect to FCI for different values of $\Omega$. Subplots (D),(E), (F)
represent the number of CNOT gates
(with parameter counts in a twin axis). The gray and red shaded areas in (D)-(F) show the range of the $\Omega$ where the energy difference lies below 0.5 $mE_h$ with respect to FCI for
AD(SPQE)-ASC (II) and SPQE, respectively. In general AD(SPQE)-ASC enters the gray region with much less number of CNOT (and parameter) counts than SPQE which represents that for achieving a similar
accuracy (in this case 0.5 $mE_h$), the former requires significantly less quantum resources due to additional ASC.
}
    \label{different_threshold_plot_equilibrium}
\end{figure*}

\subsection{Accuracy over the potential energy profile of strongly correlated chemical systems:} \label{result section 1}
In order to substantiate the effects of ASC in ideal conditions,
Fig. \ref{adspqe_pot_en_surf} shows the comparison between AD(SPQE)-ASC and
standard SPQE implementation with the
macro-iteration threshold ($\Omega$) taken to be $10^{-2}$.
In Fig. \ref{adspqe_pot_en_surf} ((A), (B), (C)) we have plotted the energy difference from Full Configuration Interaction (FCI) (in $E_h$) along the y-axis
and the ratio $(\frac{R}{R_{eq}})$ along the x-axis where, $R$ is the bond distance and $R_{eq}$ is the equilibrium distance.
For all the molecular cases AD(SPQE)-ASC (both scheme-I and II) shows better accuracy than SPQE while having the same circuit depth,
specifically with stretched geometries where the strong electronic correlations prevail.
For example in Fig. \ref{adspqe_pot_en_surf}(B) we can see that in the case of $H_6$
with $\frac{R}{R_{eq}}=2.0$ the energy error with respect to FCI for SPQE is $\sim 0.11$ $mE_h$ while that of scheme-II is $\sim 0.04$ $mE_h$,
providing almost one order-of-magnitude better energy accuracy as compared to SPQE.
The results suggest that scheme-II is the most accurate of all since it contains the additional commutator correction terms (Eq. \eqref{energy scheme II}).
The CNOT count and parameter count (for the circuits corresponding to $\hat{U}_P(\boldsymbol{\theta_P}) \ket{\Phi_0}$) are represented in Fig. \ref{adspqe_pot_en_surf} ((D), (E), (F)) 
with color codes blue and purple respectively. Here, we must re-iterate that
AD(SPQE)-ASC does not require any additional circuit resources and the CNOT gate count and number of parameters as shown in all the plots remain exactly the
same for both SPQE and our method.

However, as evident from
Fig. \ref{adspqe_pot_en_surf},
since the operator pool contains SDTQ excitations,
the CNOT counts even for SPQE at stretched geometries proliferates excessively to
account for the electronic correlation via including triples and quadruple excitation operators into the core of the ansatz.
For example, the CNOT count for $H_6$ (Fig. \ref{adspqe_pot_en_surf}(E))
reaches around $10^5$ for the stretched geometries which is way beyond the hardware capabilities of the near term quantum computers.
Thus there is a pressing need to develop methods that can attain chemically accurate description of such challenging molecular applications
with less utilization of the quantum circuit resources.
Regarding this, in subsection \ref{resource efficient implementation results} we discuss how AD(SPQE)-ASC can
provide us with a mechanism to attain chemical accuracy utilizing minimal quantum resources.

\begin{figure*}[!ht]
    \centering  
\includegraphics[width=\textwidth]{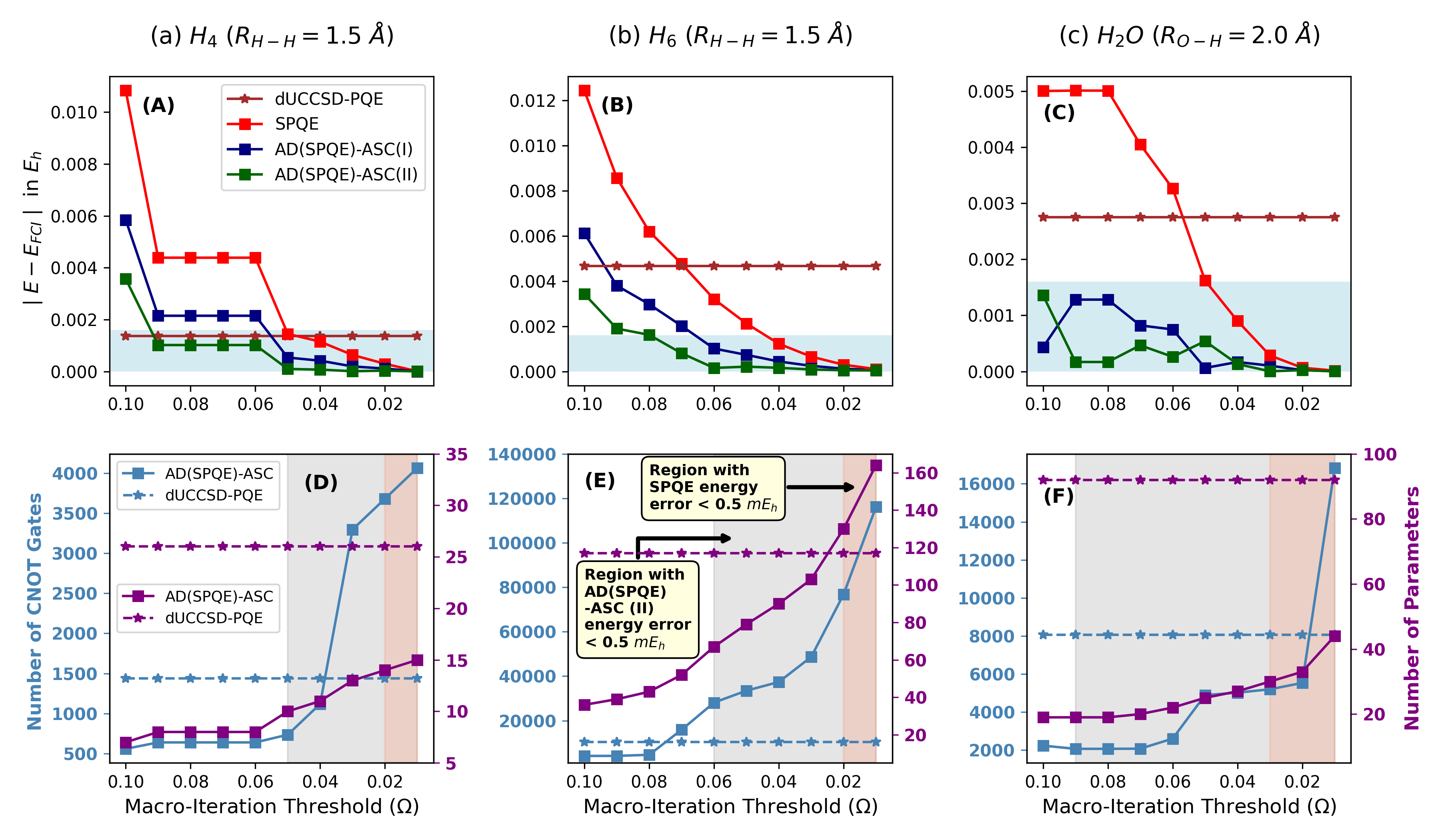}
\caption{Comparison between SPQE, AD(SPQE)-ASC (scheme-I and II) (with SDTQ pool) and dUCCSD-PQE with different macro-iteration threshold ($\Omega$) for stretched geometries of (a) $H_4$,
(b) $H_6$ and (c) $H_2O$
with the bond distances mentioned above each column.
For other details, see Fig. \ref{different_threshold_plot_equilibrium}.
}
    \label{different_threshold_plot_stretched}
\end{figure*}



\subsection{Numerical Estimation of Resource Efficiency} \label{resource efficient implementation results}

The results from section \ref{result section 1} suggest that selection of operators from SDTQ operator pool
can cause an excessive increase in circuit
depth to an impractical level considering the current NISQ devices. This can be modulated by tuning $\Omega$ to restrict the operators that are included in the core of the ansatz.
Moreover, at a lower threshold ($\Omega \sim 0.01$), AD(SPQE)-ASC and SPQE would be of similar order of accuracy in all cases as lower thresholds typically suggest
a more exhaustive inclusion of the parameters into the core of the ansatz, leaving almost no room
for a discernible improvement via ASC. Thus the most significant advantage of ASC is pronounced for the following two cases:
\subsubsection{Attaining chemical accuracy with SPQE operator pool restricted to singles and doubles:}
In order to keep the gate depth as shallow as possible, we restrict the core of the ansatz formed by only SD operators from a pool of SDTQ,
while the energy corrections due to the remaining SD and all TQ operators are embedded via ASC.
We refer to this variant as AD(SPQE-SD)-ASC to explicitly represent the operator pool type (in this case SD) considered in SPQE.
In Fig. \ref{adspqe_CCSD_TQ_pot_en_surf} we have shown the numerical comparisons between SPQE, AD(SPQE)-ASC and dUCCSD-PQE for the potential energy
surfaces of all the molecules under consideration.
The blue shaded area shows the chemical accuracy energy window ($\sim 1.6$ $mE_h$ energy error from FCI).
In case of linear $H_6$ (Fig. \ref{adspqe_CCSD_TQ_pot_en_surf}(B), (E))
scheme-II outperforms all the other methods while maintaining the chemical accuracy throughout the PES, with CNOT count as low as
$\sim$ 5000 even for stretched geometries compared to the gate count $\sim 10^5$ with SDTQ pool (Fig. \ref{adspqe_pot_en_surf}(E)).
One may note that in all these cases both dUCCSD-PQE and dUCCSD-SPQE fail to account for the strong correlation at stretched geometries. AD(SPQE-SD)-ASC
complements such incomplete inclusion of correlation via ASC to account for the missing SD and all TQ effects
and attains chemical accuracy for all the cases in our study, albeit with quantum resources equal to that of dUCCSD-SPQE.


\subsubsection{Resource Efficient Performance with higher macro-iteration threshold:}
In general, SPQE demonstrates systematically more accurate results as $\Omega$ is adjusted
from higher to lower values at the cost of increased quantum resource utilization.
In this regard, we aim to highlight the significantly improved accuracy that 
AD(SPQE)-ASC provides compared to SPQE via non-iterative ASC even at higher $\Omega$ values.
This is shown in Fig. \ref{different_threshold_plot_equilibrium} and Fig. \ref{different_threshold_plot_stretched} for
equilibrium and stretched geometries of linear $H-$chain (column (a) $H_4$ and column (b) $H_6$ 
with $R_{H-H}=0.75$ \AA, $1.5$ \AA) and $H_2O$ (column (c) with $R_{O-H}= 1.43$ \AA, $2.0$ \AA) respectively.
These comparisons between SPQE and AD(SPQE)-ASC (both schem I and II) involve varying 
$\Omega$ from 0.1 to 0.01 along the x-axis, with the energy difference from FCI
plotted in the y-axis in the first row of both Fig. \ref{different_threshold_plot_equilibrium}
and \ref{different_threshold_plot_stretched} (labeled as (A), (B) and (C)).
For reference, dUCCSD-PQE results are also shown on the same scale.
In the second row of Fig. \ref{different_threshold_plot_equilibrium}
and \ref{different_threshold_plot_stretched} ((D), (E), (F))
we have shown the CNOT count (the corresponding parameter count shown in a twin axis) 
which are, as previously discussed, same for both SPQE and AD(SPQE)-ASC.
The corresponding grey and pink shaded areas are to highlight the region where 
AD(SPQE)-ASC and SPQE energies are within 0.5 $mE_h$ from FCI, respectively.

The results from numerical simulations in Fig. \ref{different_threshold_plot_equilibrium},
\ref{different_threshold_plot_stretched} shows that for all the molecules,
AD(SPQE)-ASC (both scheme I and II) outperforms SPQE in terms of accuracy and resource efficiency.
It can be observed from the plots that with higher $\Omega$ ($\sim 0.1$), SPQE 
selects only a small number of dominant operators from the pool
which is unable to retrieve the correlation energy
and hence falls out of the chemical accuracy. On the other hand, substantial 
improvement is observed with the incorporation of ASC as it remains well within 
the chemical accuracy even with larger $\Omega$ values for all these cases.
Particularly for stretched geometries, in the case of $H_4$ and $H_6$ in
Fig. \ref{different_threshold_plot_stretched} ((D), (E)), the CNOT (parameter) count for
SPQE to enter the 0.5 $mE_h$ accuracy zone (pink shaded area) is $\sim$ 3700 (15) with
$\Omega$=0.02 and $80\times10^3$ (130) with $\Omega$=0.02, respectively. 
On the other hand AD(SPQE)-ASC enters the accuracy zone (gray region) around $\Omega=0.05$ and 0.06
with the CNOT (parameter) count $\sim$ 750 (10) and $28\times10^3$ (75) respectively,
showing a drastic reduction in the associated circuit resources. Similar trend is
observed for $H_2O$ as well as demonstrated in Figs. \ref{different_threshold_plot_equilibrium}(C, F)
and \ref{different_threshold_plot_stretched}(C, F).

As discussed earlier in section \ref{synergistic mapping}, the number of
residual element evaluations are same for SPQE and AD(SPQE)-ASC as the \textit{principal-to-auxiliary mapping}
(Eq. \eqref{theta_A as a function of theta_P}) can be in principle extracted directly from the last step of SPQE.
However, for dynamic algorithms other than SPQE,
the number of additional residual elements is $dim\{\textit{APS}\}=N_A$ at worst. The evaluation of ASC
for schemme-II requires some extra diagonal matrix elements (see last term of Eq. \eqref{energy scheme II})
which is significantly less than the cost of SPQE. Some quantitative discussions regarding the measurement cost
is discussed in Appendix \ref{measurement cost calculation}.

\section{Conclusion and Future Outlook}
In this article, we introduce a general theoretical framework
having structure like AD(x)-ASC (\enquote{x} denotes the method for preparing the core),
designed to deliver highly accurate energy calculations while significantly
reducing the quantum resource requirements to implement the dynamic ansatz.
In this particular work, we build upon SPQE algorithm, which employs an \enquote{evolve-and-measure} strategy for dynamic ansatz construction
via alternate macro and micro-iteration cycles.
We argue that SPQE naturally induces adiabatic approximation, resulting in bipartite decoupling of the operator pool.
The core (involving PPS) of the ansatz
is formed via SPQE where
only minimal number of \textit{principal} operators are chosen
through repeated quantum measurements. The remaining operators form the \textit{auxiliary} operator subspace, with parameters that can be predicted from the optimized \textit{principal} parameters using a mathematical relationship established in this work. Further, the contributions of \textit{auxiliary} operators are incorporated into the energy function via a 
set of non-iterative corrections referred to as ASC.
Thus both parts of the
ansatz are flexible, while they sum up to a predefined set of 
operator pool with fixed number of operators.
The hallmark of this entire AD(SPQE)-ASC protocol is that ASC does not require any additional circuit resources
as the circuit depth of the ansatz is explicitly governed by SPQE.
Moreover, we demonstrated that AD(SPQE)-ASC does not require any extra residual elements evaluations.

All of our numerical studies
indicate that AD(SPQE)-ASC can
substantially retrieve the electronic correlation energy
in the strongly correlated regime of the PES, even with only SD operators in the core ansatz.
It shows astounding performance specially with higher values of the operator selection threshold
where the core SPQE ansatz is not adequately accurate itself, but extremely compact.
We have also analytically and numerically substantiated that for achieving a particular order of accuracy,
AD(SPQE)-ASC requires significantly less number of CNOT gates and number of measurements as compared to SPQE.
Such exceptionally accurate performance of AD(SPQE)-ASC with
lower utilization of quantum resource requirements indicate the significance of the suite
of non-iterative corrections.
Even though we considered a projective formalism here, AD(x)-ASC is a general framework that can be coupled
with other dynamic ansatz protocols like ADAPT-VQE\cite{grimsley2019adaptive} or COMPASS\cite{mondal2023development}
for better accuracy and resource efficiency which will be pursued as separate studies.


\section{Acknowledgments}
RM acknowledges the 
financial support from Industrial Research and
Consultancy Centre (IRCC), IIT Bombay and 
Science and Engineering Research Board (SERB), Government
of India (Grant Number: MTR/2023/001306). CP  and SH acknowledges University Grants Commission (UGC)
Council of Scientific and Industrial Research (CSIR) for their respective research fellowships. DM acknowledges Prime Minister's Research Fellowship (PMRF) for his research fellowship.

\section*{AUTHOR DECLARATIONS}
\subsection*{Conflict of Interest:}
The authors have no conflict of interest to disclose.

\section*{Data Availability}
The data is available upon reasonable request to the corresponding author.

\appendix

\section{Discussion on adiabatic approximation and the determination of \textit{auxiliary} parameters from the SPQE-selected principal parameters}
\label{auxiliary mapping}

In this appendix, we discuss the adiabatic approximation and slaving principle for determination of the \textit{auxiliary} parameters from the \textit{principal} ones selected by SPQE.
With the decoupling of the parameter space, in general we can write down the parameter variation (or update) equation for both class of parameters
\begin{equation}
    \Delta \theta_{\xi} = \frac{r_\xi}{D_\xi}; \hspace{3mm} \xi \in \mbox{\textit{APS} or \textit{PPS}}
\end{equation}
where, $\xi$ is a composite hole-particle indices that can belong to \textit{principal} or \textit{auxiliary} subsets. Since the residuals $r_\xi$ are inherently nonlinear,
we can reorder the terms
\begin{equation} \label{del_tL_appendix}
\begin{split}
    \Delta \theta_{\xi} &= \Lambda_{\xi} \theta_{\xi}+ M_{\xi}(\boldsymbol{\theta_{P}},\boldsymbol{\theta_{A}}) 
\end{split}
\end{equation}
where, $\Lambda_{\xi}$ is the coefficient of the linear diagonal term
\begin{equation} \label{lambda def _appendix}
\begin{split}
    & \Lambda_{\xi} = \frac{1}{D_{\xi}} \bra{\Phi_{\xi}}  [\hat{H},\hat{\kappa}_{\xi}] \ket{\Phi_{0}}\\
\end{split}
\end{equation}
and $M_\xi$ contains all the nonlinear and linear off-diagonal terms
\begin{equation} \label{P_appendix}
    \begin{split}
        &M_{\xi}(\boldsymbol{\theta_{P}},\boldsymbol{\theta_{A}}) = \frac{1}{D_{\xi}} \bra{\Phi_{\xi}} \hat{H} + \sum_{\nu \neq \xi} \theta_{\nu} [\hat{H},\hat{\kappa}_{\nu}] \\
 & + \sum_{\nu} \sum_{\mu} \theta_{\nu}
 \theta_{\mu} \Big[[\hat{H},\hat{\kappa}_{\nu}],\hat{\kappa}_{\mu}\Big] + . . . \ket{\Phi_0}
    \end{split}
\end{equation}
Due to the relatively slower convergence of the \textit{principal} parameters, the
variation of the \textit{auxiliary} parameters can be neglected in the characteristic convergence timescale of the \textit{principal} parameters.
This is known as the \textit{adiabatic approximation} \cite{patra2024projective,patra2023synergistic,agarawal2021approximate}.
Due to this, we can set \textit{auxiliary} variation $\Delta \theta_{A_\alpha}=0$ in the update equation Eq. \eqref{del_tL_appendix}
\begin{equation} \label{intermediate auxiliary eqn}
    0 = \Lambda_{A_\alpha} \theta_{A_\alpha}+ M_{A_\alpha}(\boldsymbol{\theta_{P}},\boldsymbol{\theta_{A}})
\end{equation}
With this we can re-order Eq. \eqref{intermediate auxiliary eqn} for the adiabatic solution of the \textit{auxiliary} parameters
\begin{equation} \label{ts_gen_appendix}
    \theta_{A_\alpha} = -\frac{M_{A_\alpha}(\boldsymbol{\theta_{P}},\boldsymbol{\theta_{A}})}{\Lambda_{A_\alpha}} 
\end{equation}
Further, we can invoke the relative magnitude condition (Eq. \eqref{parameter amplitude condition}) to express the \textit{auxiliary} parameters as a function of \textit{principal} parameters only
\begin{equation} \label{thetaS_ad_appendix_1}
    \begin{split}
        \theta_{A_\alpha} &= -\frac{M_{A_\alpha}(\boldsymbol{\theta_{P}},\boldsymbol{\theta_{A}})}{\Lambda_{A_\alpha}} \xrightarrow{{\mid \theta_A \mid } << {\mid \theta_P \mid }} \approx {-\frac{M_{A_\alpha}(\boldsymbol{\theta_{P}})}{\Lambda_{A_\alpha}}}\\
    \end{split}
\end{equation}
From Eq. \eqref{P_appendix} one can write $M_{A_\alpha}(\boldsymbol{\theta_{P}})$ explicitly as 
\begin{equation} \label{M_as_func_of_thetaP}
    \begin{split}
        &M_{A_\alpha}(\boldsymbol{\theta_{P}}) \approx \frac{1}{D_{A_\alpha}} \bra{\Phi_{A_\alpha}} \hat{H} + \sum_{P_I} \theta_{P_I} [\hat{H},\hat{\kappa}_{P_I}] \\
 & + \sum_{P_I} \sum_{P_J} \theta_{P_I}
 \theta_{P_J} \Big[[\hat{H},\hat{\kappa}_{P_I}],\hat{\kappa}_{P_J}\Big] + . . . \ket{\Phi_0}\\
        & = \frac{\bra{\Phi_{A_\alpha}} \hat{U}_{P}^{\dagger}(\boldsymbol{\theta_{P}}) \hat{H} \hat{U}_{P}(\boldsymbol{\theta_{P}}) \ket{\Phi_0}}{D_{A_\alpha}}
    \end{split}
\end{equation}
Using the expression obtained in Eq. \eqref{M_as_func_of_thetaP}, Eq. \eqref{thetaS_ad_appendix_1} can be compactly written as-
\begin{equation} \label{thetaS_ad_appendix}
    \begin{split}
        \theta_{A_\alpha} &= -\frac{\bra{\Phi_{A_\alpha}} \hat{U}_{P}^{\dagger}(\boldsymbol{\theta_{P}}) \hat{H} \hat{U}_{P}(\boldsymbol{\theta_{P}}) \ket{\Phi_0}}{\bra{\Phi_{A_\alpha}}  [\hat{H},\hat{\kappa}_{A_\alpha}] \ket{\Phi_{0}}} \\
        & = \frac{\bra{\Phi_{A_\alpha}} \hat{U}_{P}^{\dagger}(\boldsymbol{\theta_{P}}) \hat{H} \hat{U}_{P}(\boldsymbol{\theta_{P}}) \ket{\Phi_0}}{D_{A_\alpha}}
    \end{split}
\end{equation}
where,
${\bra{\Phi_{A_\alpha}}  [\hat{H},\hat{\kappa}_{A_\alpha}] \ket{\Phi_{0}}} \approx -D_{A_\alpha}$ under leading order
approximation (see Appendix \eqref{app:section1}).

\section{First Order Approximation of Commutator Expectation Values}
\label{app:section1}

In this section, we provide a mathematical details of the approximation of the denominator for \textit{auxiliary} solution Eq. \eqref{thetaS_ad_appendix}.
The commutator $\bra{\Phi_{\mu}}  [\hat{H},\hat{\kappa}_{\nu}] \ket{\Phi_{0}}$
can be approximated by expanding the Hamiltonian into a zeroth-order one-body Fock operator and a first-order two-body operator,
$\hat{H}=\hat{F}^{(0)} + \hat{V}^{(1)}$.
Retaining only the zeroth-order term we get
\begin{equation} \label{commutator apprximation appendix}
\begin{split}
    & \bra{\Phi_{\mu}}  [\hat{H},\hat{\kappa}_{\nu}] \ket{\Phi_{0}}  \\
    \implies & \bra{\Phi_{\mu}}  [\hat{F}^{(0)},\hat{\kappa}_{\nu}] \ket{\Phi_{0}} \approx -D_{\mu} \delta_{\mu \nu}
\end{split}
\end{equation}



\section{Derivation of the low depth energy determining equation with non-iterative corrections}
\label{appendix: energy equation derivation}
Here, we analytically establish the energy expression for scheme-I (Eq. \eqref{energy scheme I}) and scheme-II (Eq. \eqref{energy scheme II}) for a shallow-depth
inclusion of the \textit{auxiliary} parameters into the energy functional.
With the \textit{principal-auxiliary bipartite} operator (Eq. \eqref{U_pab principal auxiliary bipartite op}), the energy term can be expressed by 
\begin{equation} \label{energy with feedback}
   E(\boldsymbol{\theta}) = \bra{\Phi_o}\hat{U}^{\dagger}_{pab}(\boldsymbol{\theta})H \hat{U}_{pab}(\boldsymbol{\theta})\ket{\Phi_o}
\end{equation}


\textit{Term 1} and \textit{Term 2} in Eq. \eqref{energy expression} can be further expanded as follows:

\subsubsection*{Term 1}
\begin{equation} \label{Energy derivation Term 1}
\begin{split}
    & \theta_{A_\alpha} \bra{\Phi_{0}}  [\bar{H}_P,\hat{\kappa}_{A_\alpha}] \ket{\Phi_{0}} \\
    & =  \theta_{A_\alpha} \bra{\Phi_{0}} \Bar{H}_P \hat{\kappa}_{A_\alpha} \ket{\Phi_0} - \bra{\Phi_{0}} \hat{\kappa}_{A_\alpha} \Bar{H}_P  \ket{\Phi_0} \\
    & =  \theta_{A_\alpha} \bra{\Phi_{0}} \Bar{H}_P \ket{\Phi_{A_\alpha}} + \bra{\Phi_{A_\alpha}}  \Bar{H}_P  \ket{\Phi_0} \\
    & =  2 \theta_{A_\alpha} \bra{\Phi_{A_\alpha}} \hat{U}_P^{\dagger} \hat{H} \hat{U}_P \ket{\Phi_0} \\
    & = 2 \theta_{A_\alpha}^2 D_{A_\alpha}
\end{split}
\end{equation}
where, at the last step, we have used Eq. \eqref{theta_A as a function of theta_P} to replace the operator expectation value with adiabatically obtained \textit{auxiliary} parameters.
\begin{figure*}[!ht]
    \centering  
\includegraphics[width=\textwidth]{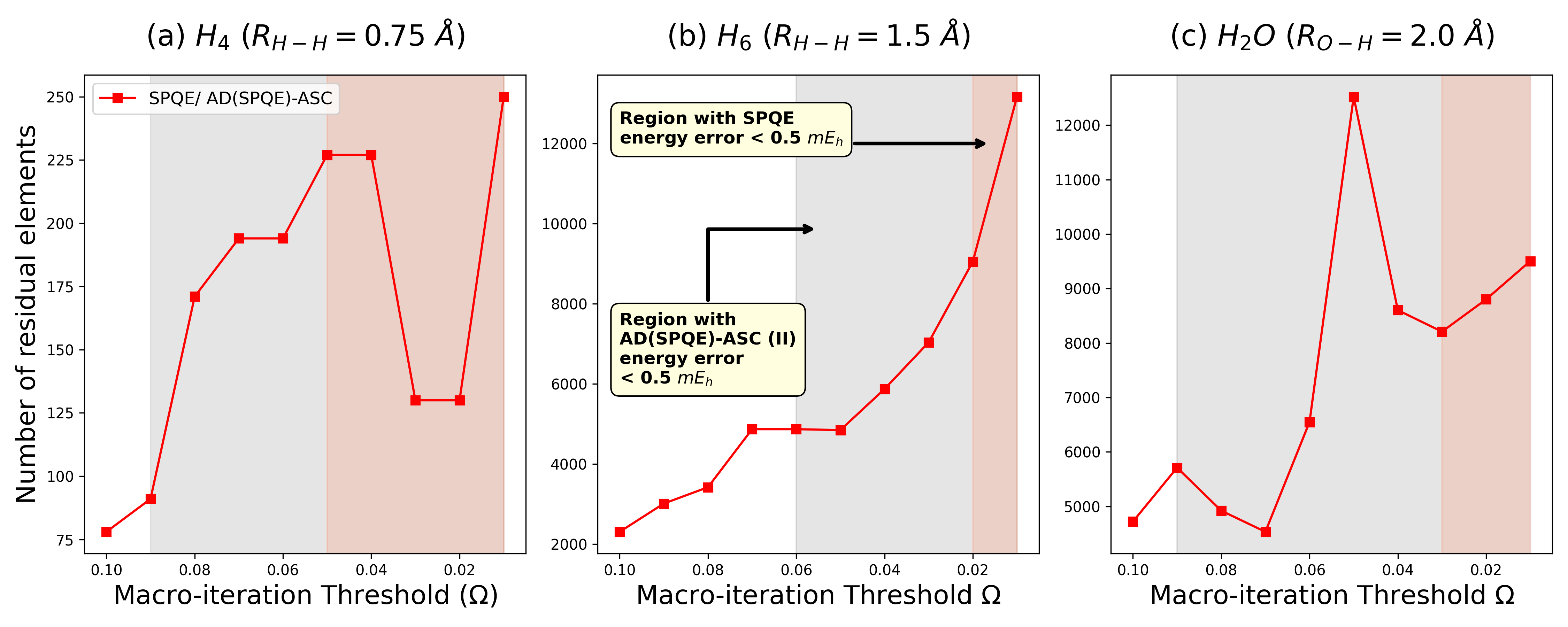}
\caption{Number of residue element evaluations for SPQE and AD(SPQE)-ASC (with SDTQ pool) with different macro-iteration threshold ($\Omega$). Note that, the number of residue elements
for both the methods are same for a particular value of $\Omega$. However, for attaining similar order of accuracy, AD(SPQE)-ASC (II) requires less number of residue elements than SPQE as respresented by different shaded regions (see the subplot corresponding to column (b) for the details of the color coding).
}
    \label{residue_count_plot}
\end{figure*}
\subsubsection*{Term 2}
Since \textit{Term 2} is a nonlinear term, for its evaluation, we first approximate the BCH expansion for $\Bar{H}_P$ up to zeroth order,
which is just the bare Hamiltonian
\begin{equation} \label{term 2 hamiltonian approx}
\begin{split} %
    & \bra{\Phi_{0}} \Big[[\Bar{H}_P,\hat{\kappa}_{A_\alpha}],\hat{\kappa}_{A_\beta}\Big] \ket{\Phi_{0}} \approx \bra{\Phi_{0}} \Big[[\hat{H},\hat{\kappa}_{A_\alpha}],\hat{\kappa}_{A_\beta}\Big] \ket{\Phi_{0}}\\
\end{split}
\end{equation}
For scheme-I, we can approximate the Hamiltonian to the zeroth order fock operator only for a simpler expression
\begin{equation} \label{Energy derivation Term 2  scheme I}
\begin{split}
    & \frac{1}{2}\theta_{A_\alpha} \theta_{A_\beta} \bra{\Phi_{0}} \Big[[\Bar{H}_P,\hat{\kappa}_{A_\alpha}],\hat{\kappa}_{A_\beta}\Big] \ket{\Phi_{0}} \\
    & \approx  \frac{1}{2}\theta_{A_\alpha} \theta_{A_\beta} \bra{\Phi_{0}} \Big[[\hat{H},\hat{\kappa}_{A_\alpha}],\hat{\kappa}_{A_\beta}\Big] \ket{\Phi_{0}}\\
    & \approx  \frac{1}{2}\theta_{A_\alpha} \theta_{A_\beta} \Big( \bra{\Phi_{0}}[\hat{F}^{(0)},\hat{\kappa}_{A_\alpha}] \hat{\kappa}_{A_\beta} \ket{\Phi_{0}} - \\ 
    & \bra{\Phi_{0}} \hat{\kappa}_{A_\beta} [ \hat{F}^{(0)},\hat{\kappa}_{A_\alpha}] \ket{\Phi_{0}}\Big) \\
    & =  \frac{1}{2}\theta_{A_\alpha} \theta_{A_\beta} \Big( \bra{\Phi_{0}}[\hat{F}^{(0)},\hat{\kappa}_{A_\alpha}] \ket{\Phi_{A_\beta}}+ \\
    & \bra{\Phi_{A_\beta}} [ \hat{F}^{(0)},\hat{\kappa}_{A_\alpha}] \ket{\Phi_{0}}\Big) \\
    & =  \theta_{A_\alpha} \theta_{A_\beta} \bra{\Phi_{A_\beta}} [ \hat{F}^{(0)},\hat{\kappa}_{A_\alpha}] \ket{\Phi_{0}} \\
    & \approx -  \theta_{A_\alpha} \theta_{A_\beta} D_{A_\beta} \delta_{A_\alpha A_\beta} 
\end{split}
\end{equation}
where, at the last step we have used the expression in Eq. \eqref{commutator apprximation appendix}.
Plugging the final forms of \textit{Term 1} (Eq. \eqref{Energy derivation Term 1}) and \textit{Term 2} (Eq. \eqref{Energy derivation Term 2  scheme I})  in Eq. \eqref{energy expression} we get
the scheme-I energy expression Eq. \eqref{energy scheme I}.
In Eq. \eqref{Energy derivation Term 2  scheme I} the contributions from the off-diagonal terms are zero due to the presence of the delta function.
Even in the general case, the contributions from the off-diagonal terms are negligible such that Eq. \eqref{term 2 hamiltonian approx} can be assumed to contain
only diagonal terms which leads to the following expression
\begin{equation} \label{term 2 hamiltonian diagonal approx}
\begin{split} 
    & \bra{\Phi_{0}} \Big[[\Bar{H}_P,\hat{\kappa}_{A_\alpha}],\hat{\kappa}_{A_\beta}\Big] \ket{\Phi_{0}} \approx \bra{\Phi_{0}} \Big[[\hat{H},\hat{\kappa}_{A_\alpha}],\hat{\kappa}_{A_\alpha}\Big] \ket{\Phi_{0}}\\
\end{split}
\end{equation}
Plugging this along with the expression obtained in Eq. \eqref{Energy derivation Term 1} into energy expression Eq. \eqref{energy expression}
results in the scheme-II energy equation Eq. \eqref{energy scheme II} which is more accurate than scheme-I as discussed in section \ref{results and discussion section}.

\section{Measurement overhead for auxiliary subspace correction terms} \label{measurement cost calculation}
The number of measurements in SPQE for all the micro-iterations is
of the order of\cite{stair2021simulating}
\begin{equation} \label{measurement cost of SPQE}
    M_{SPQE}  \leq \hspace{3mm} N_{res}\times3N_P\frac{(\Sigma_l h_l)^2}{\epsilon^2}
\end{equation}
while, the measurement overhead required to converge SPQE macro-iterations with threshold $\Omega$ is
$\sim (\Delta t \Omega)^{-2} $.
Here, $N_{res}$ is the total number of residual vectors evaluated during all of the micro-iteration optimization,
$h_l$ is the coefficient of $l$-th Pauli
string in the Jordan-Wigner mapped Hamiltonian and $\epsilon$ is the desired precision in measurement based
energy estimation. Since the mapped \textit{auxiliary} parameters via Eq. \eqref{theta_A as a function of theta_P}
can directly be obtained from SPQE-excluded residuals, scheme-I does not require any additional measurements.

The additional measurement for ASC in scheme-II due to the third term of Eq. \eqref{energy scheme II} is
\begin{equation} \label{scheme-II measurement cost}
    M_{II} \leq N_A \frac{(\Sigma_l h_l)^2}{\epsilon^2}
\end{equation}
This additional measurement cost $M_{II}$
stems from $\bra{\Phi_0} * \ket{\Phi_0}$ type of measurements in Eq. \eqref{energy scheme II}.
Thus the total measurement cost of AD(SPQE)-ASC (II) becomes
\begin{equation} \label{adspqe_asc measurements}
    M_{AD(SPQE)-ASC} = M_{SPQE} + M_{II}
\end{equation}
For scheme-II, the additional cost for ASC ($M_{II}$) is in general less than that of
SPQE ($M_{SPQE}$) as $N_A$ is usually less than $( N_{res} \times 3 N_P )$ for all practical scenarios.
Note that, both the schemes of AD(SPQE)-ASC
do not require any extra residual elements to be evaluated in addition to that of SPQE.
In Fig. \ref{residue_count_plot} we have shown the number of residue element evaluations for
different $\Omega$ values. Note that the curve enters the gray region with much less number of residue element count
which represents that for attaining similar order of accuracy (in this case $<0.5$ $mE_h$), AD(SPQE)-ASC (II) requires much less number
of residue elements (and measurements) to be evaluated than standard SPQE.

\section*{References:}


%

\end{document}